\documentclass[twocolumn,showpacs,preprintnumbers,amsmath,amssymb,aps]{revtex4}
\usepackage{graphicx}
\usepackage{dcolumn}

\usepackage{color}

\newcommand{\be}{\begin{equation}}
\newcommand{\ee}{\end{equation}}
\newcommand{\bea}{\begin{eqnarray}}
\newcommand{\eea}{\end{eqnarray}}
\newcommand{\ds}{\displaystyle}

\begin{document}
\title{
Current  limiting effects on laser compression by resonant backward Raman scattering}
\author{Nikolai~A.~Yampolsky}
\affiliation{Los Alamos National Laboratory, Los Alamos, New Mexico, 87545, USA}
\author{Nathaniel~J.~Fisch}
\affiliation{Department of Astrophysical Sciences, Princeton University, Princeton, New Jersey 08544, USA}


\begin{abstract}
{Through resonant backward Raman scattering, the plasma wave mediates the energy transfer between long pump and short seed laser pulses.  These mediations can result in  pulse compression at extraordinarily high powers.  However, both the overall efficiency of the energy transfer and the duration of the amplified pulse depend upon the persistence of the plasma wave excitation. At least with respect to the recent state-of-the-art experiments, it is possible to deduce that  at present the experimentally realized efficiency of the amplifier is likely constrained mainly by two effects, namely the pump chirp and the plasma wave wavebreaking.  }
\end{abstract}


\maketitle


\section{Introduction}
High power femtosecond laser pulses are currently generated through the chirped pulse
amplification (CPA) scheme \cite{CPA}.   The damaging of the solid state amplifier by
high power radiation is avoided by the temporal stretching of the laser pulse, its amplification to high
intensity, and only then its compression.  However, the high power radiation can still damage solid state
elements of the CPA laser at the compression stage.  Thus, the peak intensity and 
fluence at the last compressing grating should remain below the critical values, so that higher
overall CPA laser power can be reached only if larger size gratings are used. That significantly
increases the cost of the CPA laser and CPA technology approaches its practical limit.

Alternatively, plasma can be used as a nonlinear medium to mediate laser amplification, since
it cannot be damaged by high power radiation. The resonant backward Raman amplifier \cite{prl99,BRA_overview}
is at present the most promising and the best studied plasma-based scheme for high power laser
amplification. This scheme relies on the resonant coupling of two counterpropagating laser waves
mediated by the plasma wave.
The counterpropagating geometry in BRA allows one to convert  the long pump
energy into the energy of the short seed pulse 
.
The resonant nature of interaction allows one to use a relatively low intensity pump pulse and
reach significantly higher efficiencies compared to mechanisms based on nonresonant wave coupling
\cite{superradiant}. 

The resonant conditions for the wave coupling are
\be
\label{resonant}
\omega_a=\omega_b+\omega_p,\;\;\;\;\;\;\;\;\;\;\;\;\;
k_a=k_b+k_p,
\ee
where $\omega_a$, $\omega_b$, and $\omega_p=(4\pi e^2n/m)^{1/2}$
are the frequencies of the pump, seed, and plasma
wave, respectively; $k_a$, $k_b$, and $k_p$ are the wavenumbers of the pump, seed, and the
plasma wave, respectively. These conditions are easy to satisfy since the plasma wave dispersion
relation weakly depends on the plasma wavenumber and so one needs just to choose appropriate plasma
density $n$ for efficient laser coupling.

The BRA can be considered as a photon decay instability of the pump photon into the seed photon and
the plasmon. Resonant conditions (\ref{resonant}) can be interpreted then as the energy and
momentum conservation laws of the decay. The pump
energy transferred into the plasma wave energy can be small as long as the plasma-to-laser
frequency ratio remains small.  However, the plasma density
should be high enough to provide significant laser coupling.   In the nonlinear stage, in which there is  complete pump depletion,
the BRA efficiency approaches the theoretical limit of $1-\omega_p/\omega_a\approx 1$.

This paper will review the implications of the current plasma-based resonant BRA experiments,  including those at  performed at Princeton
University \cite{Suckewer00,Suckewer03,Suckewer04,Suckewer05,Suckewer07}, Institute of Applied
Physics (IAP) in Russia \cite{IAP_exp}, Lawrence Livermore National Laboratory (LLNL)
\cite{LLNL07,LLNL09,LLNL11}, and Institute of Atomic and Molecular Sciences (IAMS) in Taiwan \cite{Tai_exp}.
Although by some measures still large, the  experimentally achieved efficiencies were significantly lower
than 
theoretical limits. Earlier analytical 
\cite{PoP00,Tsidulko_scat,precursor,Dodin_cap,Solodov_dens,Solodov_side,Clark_Solodov,high_detune,Malkin_quasi1}
and numerical \cite{Solodov_dens,Clark_Solodov,robustness,Mardahl_cap,self_ion_num,Balakin_multipump,Clark_PIC,Balakin_ion,Hur06}
studies were focused on identifying mechanisms which may limit the BRA efficiency and finding regimes where these
limitations can be avoided \cite{Clark_q,Wang_PIC,Malkin_quasi2,PIC_Nature,Farmer10}.   It is of interest, therefore, to assess what are the current limitations in
experimental achievements.  

Although these experiments are only partially diagnosed, it is nonetheless possible to rule out some
limiting mechanisms.  For example, none of the experiments reported
significant signal corresponding to the forward Raman scattering (FRS) of the amplified pulse,
unlike the concern raised in several numerical studies \cite{Wang_PIC,PIC_Nature}. Also, the efficiency in 
high power LLNL experiment was similar to the efficiencies in other small scale experiments,
which indicates that self-focusing and self-phase modulation instabilities \cite{prl99,PoP00,robustness,PIC_Nature}
do not affect amplification in {\it current} experiments. 
However, the influence of other limiting effects is less obvious. It is the purpose of this paper to identify these limiting effects in {\it current}
BRA experiments, thereby to improve upon these experiments by avoiding the current limitations.

The paper is organized as Follows:  In Sec.~\ref{basic}, we introduce the basic model describing wave coupling in BRA. In Sec.~\ref{detuning} we demonstrate
that detuning can significantly reduce the BRA efficiency in current experiments, but that the detuning caused by the pump chirp can be compensated by the plasma
density gradient. In Sec.~\ref{wavebreaking} we discuss the influence of plasma wavebreaking on pulse amplification. In Sec.~\ref{window} we demonstrate that the combined
limitation due to detuning and the wavebreaking suggests a pump intensity window for efficient BRA. In Sec.~\ref{Landau} we estimate the highest plasma temperature that may be tolerated in view of the increased Landau damping of the plasma wave. In Sec.~\ref{epsilon} we discuss the requirements on the seed pulse quality.


\section{Basic model}
\label{basic}

The parameters of the above experiments (except for the experiment
at IAP)  were similar to each other: amplification was performed by a pump with 
intensities on the order of $I_a\sim10^{14}\, {\rm W/cm^2}$ in a few millimeter plasma having density $n\sim10^{19}\,  {\rm cm^{-3}}$.
These parameters are close to those suggested for BRA
experiments  \cite{PoP00}. The plasma-to-laser frequency
ratio under these conditions is on the order of $1/10$, which enables strong laser coupling
on one hand, but keeps high maximum laser conversion efficiency (on the order of $90 \%$),
on the other hand. 
In these regimes, neglecting focusing and refraction effects, the BRA 
can be described adequately in terms of a set of equations for the wave envelopes of circularly polarized laser pulses \cite{prl99,PoP00},
\bea
\label{eqa}
&&a_t+ca_z=-\sqrt{\omega\omega_p}bf\\
&&b_t-cb_z=\sqrt{\omega\omega_p}af^*\\
\label{eqf}
&&f_t-i\delta\omega f=\sqrt{\omega\omega_p}ab^*/2,
\eea
where $a, b$ are the amplitudes of the pump and the seed, respectively, normalized so that the laser intensity
$I_{a}=\pi c(mc^2/e)^2|a|^2/\lambda^2=2.736\times10^{18}|a|^2/\lambda^2[\mu m]$ W/cm$^2$; $f$ is appropriately normalized
amplitude of the plasma wave; $\delta\omega$ is the detuning frequency between the waves which
is caused by failure of exact resonant condition (\ref{resonant}); $t$ is time and
$z$ is longitudinal coordinate along the direction of the pump propagation.

The advanced stage of amplification is characterized by pump depletion,  
where Eqs.~(\ref{eqa}) --- (\ref{eqf}) reduce to 
\be
\label{eqa_reduced}
a_\zeta=-bf,\;\;\;\;b_\tau=af^*,\;\;\;\;f_\zeta-ia_0^2q\tau f=ab^*,
\ee
where $\zeta=(t+z/c)\sqrt{\omega_a\omega_p}/2$ is the normalized longitudinal coordinate behind the seed pulse and
$\tau=-z\sqrt{\omega_a\omega_p}/c$ is the normalized amplification length,
$q=2(\partial_z\omega_p-2\partial_z\omega_a)c/(a_0^2\omega_a\omega_p)$ is the detuning factor \cite{PoP00}. 

This set of equations allows a self-similar solution \cite{PoP00} having the following form
\bea
\label{self1}
&b=a_0^2\tau B(a_0^2\zeta\tau)&\\
\label{self2}
&a=a_0 A(a_0^2\zeta\tau),\;\;\;\;\;\;\; f=a_0F(a_0^2\zeta\tau)&.
\eea
In this regime the maximum amplitude of the amplified pulse, $b_{\rm max}(\tau)$,
grows linearly with the amplification length while its duration decreases
inversely proportional with the amplification length; $b_{max}\propto \tau$, $\Delta\zeta\propto1/\tau$. Moreover, at small values of detuning, $q\ll1$, the pump is almost completely depleted
after interacting with the amplified pulse \cite{PoP00}.  The efficiency of BRA then approaches $1-\omega_p/\omega$.
These scalings were observed experimentally \cite{Suckewer05}, where  it was concluded that the amplification reached the nonlinear stage accompanied by
significant pump depletion. Similar features were also observed in recent experiments  \cite{LLNL09,LLNL11}, in which
the output pulse energy scales linearly with the pump energy and is almost independent of the seed pulse energy.

However, these features do not necessarily prove that significant pump depletion was actually
reached. The same scalings of the amplified pulse energy can be described by Eqs.~(\ref{eqa_reduced}) with significant detuning
$q\sim1$ \cite{PoP00,high_detune}.   Thus, measurements of the output pulse energy scalings can be a useful tool suggesting 
different regimes of BRA but these scalings cannot determine without doubt whether significant pump depletion was
experimentally reached. Only accurate estimate of the pump depletion within the temporal and spatial overlap of the pump and
the amplified pulses can be a reliable tool for determining whether amplification reached the nonlinear stage. 

%

The latest experiments
with tilted laser geometry at Princeton \cite{Suckewer07} can be deduced to reach significant pump depletion, perhaps
about 0.4 \cite{demonstration}. The pump depletion in other experiments appears to be
well below this level ($\sim0.1\%$ in IAP
experiment, $\sim0.3\%$ in LLNL experiments, and $\sim0.5\%$ in IAMS experiment). These estimates
suggest that nonlinearity caused by the pump depletion is not the leading limiting mechnism in current experiments and the observed scalings are caused by 
other physical effects. In the following sections we describe the mechanisms which are likely to play significant role in current experiments limiting the BRA efficiency.

\section{Detuning}
\label{detuning}

The BRA mechanism relies on the resonant interaction between waves. Deviations from the
resonance condition (\ref{resonant}) change the instability bandwidth and may significantly
reduce wave coupling. This reduction is the largest when the instability bandwidth constantly
changes over the amplification length. Often the shift of the amplified pulse 
resonant frequency can be approximated with a linear function and BRA can be described by
Eqs.~(\ref{eqa_reduced}).
The detuning can be characterized by the dimensionless parameter
\be
\label{q}
q=\frac{2(\partial_z\omega_p-2\partial_z\omega_a)c}{a_0^2\omega_a\omega_p},
\ee
where $a_0$ is the undepleted amplitude of the pump. It describes the change of the detuning frequency in the frame moving
with the amplified pulse $\delta\omega\propto q(t+z)$ in Eq.~(\ref{eqf}). Detuning results in incomplete pump depletion, which, in turn,  limits the BRA efficiency.
For an infinitely short seed pulse having integrated amplitude $\epsilon=\sqrt{\omega_a\omega_p}\int b_0 dt/2$, the pump depletion rate can be approximated by an asymptotic expression in the limit
$q,\epsilon \ll1$ \cite{PoP00}
\be
\label{depletion_detuning}
D_{a}\approx1-q^2\frac{\xi_*^2}{16},
\ee

There are two primary mechanisms which can drive the interaction away from the exact
resonance. The detuning can be caused either by the plasma density gradient or by the pump chirp. Early experiments at
LLNL \cite{LLNL07} suggest that spontaneous Raman spectrum
broadening can be explained by plasma density fluctuations produced by a speckled pump. Random detuning caused by these density fluctuations can strongly reduce amplification.
This scenario can be determined only upon availability of detailed plasma density
measurements which will determine the characteristic gradient for the density variations. Alternatively, the detuning can be caused by the pump chirp.
Many current BRA experiments are pumped with CPA laser systems \cite{Suckewer03,Suckewer04,Suckewer05,Suckewer07,IAP_exp,Tai_exp}, which are characterized by a large
bandwidth. When the CPA laser pulse is stretched, the induced pulse chirp can be large enough resulting in significant detuning parameter $q_a$.
\be
\label{domega}
q_a\approx \frac{613}{\Delta t_a[{\rm ps}]}\frac{\Delta\omega_a}{\omega_a}\left(\frac{10^{19}{\rm cm^{-3}}}{n}\right)^{1/2}\frac{10^{14}{\rm W/cm^2}}{I_a \lambda^2[{\rm \mu m}]},
\ee
where $\Delta t_a$ is the FWHM pump duration and $\Delta\omega_a/\omega_a$ is the relative pump bandwidth. In the experiments performed at Princeton, IAP, and IAMS,
the pump intensity and bandwidth
were similar to each other. However, the detuning parameter induced by the pump chirp was significantly
different. 

The importance of the pump chirp in BRA experiments was
recognized first in the experiment performed at IAP \cite{IAP_exp} 
which was identified as one of the main causes of limited amplification. 
The large values of detuning parameter in that experiment, $q\approx1$, result mainly from low plasma density,
$n=(1.4-4)\times10^{16}$ cm$^{-3}$. 
On the other hand, detuning was less important in other experiments. 
The plasma density in IAMS experiment \cite{Tai_exp} was roughly two orders of magnitude higher than in IAP experiments.
Under these conditions, the detuning in IAMS experiments was on the order of $q\sim0.1$ and
most likely did not significantly affect pulse amplification. However, the detuning factor can be large even in dense plasma at small pump pulse duration similar to
experiments performed at Princeton University \cite{Suckewer07,demonstration}. In that setup the pump duration was about 20 ps, roughly equal to the seed pulse
amplification time in 2 mm plasma resulting in $q\approx0.27$.

The detuning parameter caused by the pump chirp can be reduced either by increasing the pump duration or 
intensity as follows from Eq.~(\ref{domega}). These methods are not efficient for increasing the BRA efficiency since
high pump intensity can result in the plasma wavebreaking as will be discussed in Sec.~\ref{wavebreaking} and increased pump duration results in seed amplification by only a fraction of the pump.
Alternatively, the detuning caused by the pump chirp can be partially compensated by detuning caused by the plasma density gradient as follows from Eq.~(\ref{q})
In this regime, the resonant frequency for the seed amplification remains constant in the frame moving with the seed pulse, {\it i.e.} $(\partial_t-c\partial_z)(\omega_a-\omega_p)=0.$ 
\\

\begin{figure}[ht]
\includegraphics[width=3.5in,keepaspectratio]{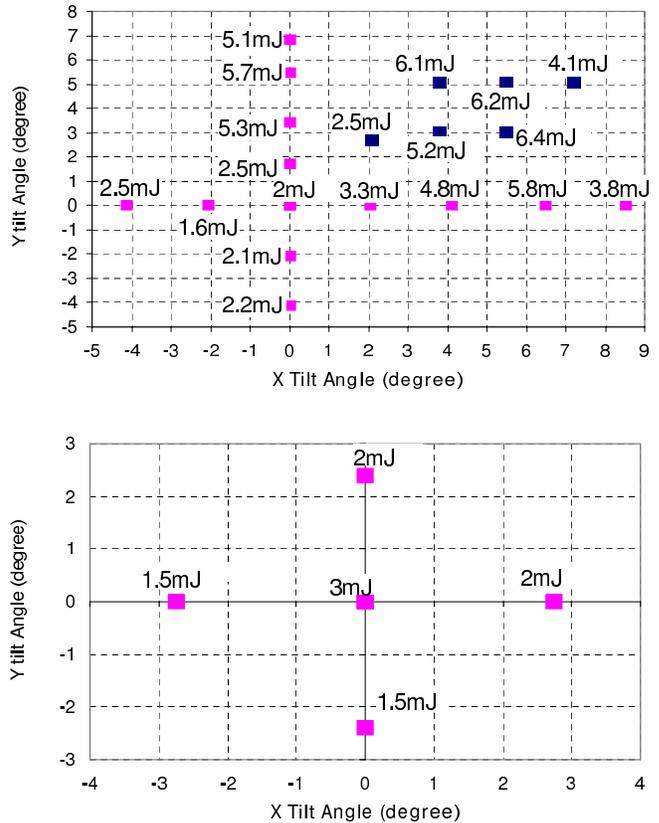}
\caption{(Color online)
The output energy of the amplified pulse versus the tilt angle of the laser beams. Experiments with positive chirp (upper plot) demonstrate increased BRA efficiency
in tilted geometry unlike negative chirp experiments (lower plot) which demonstrate the maximum BRA efficiency at axial propagation of the laser pulses. The parameters of the experiment
are described in Ref.\cite{demonstration}.}
\label{f3}
\end{figure}

This was first illustrated \cite{demonstration}  following Princeton experiments \cite{Suckewer07},
in which two laser pulses propagated at an angle with respect to the channel axis.  In this setup,
the plasma frequency changes along the interaction path since the channel has a transverse density gradient.
The tilted geometry of laser propagation resulted in about a factor of 3 larger amplified pulse energy compared to axial experiments
(upper plot of Fig.~\ref{f3}).
It is critical to note that the enhanced pump depletion in experiments
with tilted alignment of the laser beams can be observed only if the detunings have opposite signs and partially cancel each other. 

The detuning compensation was demonstrated by changing the sign of the pump chirp, as seen in Fig.~\ref{f3} \cite{demonstration}.  It is noteworthy that this demonstration is conclusive even without an accurate estimate of the plasma density gradient in the interaction region.
In the experiments with negative (reversed) pump chirp (lower plot),
the efficiency in tilted experiments is smaller than the efficiency in axial
experiments. This tendency is opposite to the experiments with positive pump chirp (upper plot).
This data verifies that the sign of the pump chirp is very important for BRA efficiency.
Flipping the sign of the pump chirp results in the addition of the
detunings rather than their subtraction.  The overall detuning factor $q$ increases so that  the efficiency drops
significantly.

The sensitivity of the overall detuning factor to the sign of the pump chirp should be taken
into account while planning future experiments.
Choosing the wrong chirp sign (positive or negative depending on the sign of the plasma density gradient) can result in increased values of the detuning factor and reduce the BRA
efficiency similar to IAP experiments \cite{IAP_exp} in which the plasma density increased and the pump frequency decreased in the frame moving with the seed pulse (Fig.~5 of Ref.~\cite{IAP_exp}).

\section{Plasma wavebreaking}
\label{wavebreaking}

Another mechanism which can prevent full pump depletion is plasma wavebreaking which occurs
when the plasma wave is too strong.   When the plasma wave grows, the electron oscillations in the plasma wave field also grow.
Electrons originating at different spatial positions then oscillate with different phases, since
the phase of the plasma wave changes longitudinally, $E_p\propto\exp(ik_pz)$.
The electron flow is not laminar then, damping the  plasma wave. 


The amplified pulse energy at the highest pump intensity is almost independent of the input pulse energy in several BRA experiments  \cite{LLNL07,IAP_exp, Suckewer05}.
Since the plasma wave at constant pump depletion is proportional to the pump intensity, this suggests that amplification may be limited  in its nonlinear stage  by the plasma wavebreaking. 
Assuming that the pump can be fully depleted in BRA, one can find the critical pump intensity $I_a^{cr}$
which leads to the plasma wavebreaking \cite{PoP00,prl99}
\be
\label{Icrit}
I_{a}^{cr}=1.45\cdot10^{14}\lambda[{\rm \mu m}] \left(\frac{n}{10^{19}{\rm cm}^{-3}}\right)^{3/2} {\rm W/cm^2}.
\ee
However, finite plasma temperature can significantly reduce this threshold \cite{Coffey,Wang_PIC}.

The plasma wavebreaking effect is not analytically tractable.
Some heuristic models for the plasma wavebreaking were used  to explain results of IAP experiment \cite{IAP_exp}.
The experimental data might fit the so-called ``soft'' wavebreaking model in which the amplitude of the plasma wave remains  constant after reaching the critical value \cite{IAP_exp}.
This model assumes that the plasma wave does not vanish after
reaching critical amplitude so that the pulse amplification survives in the wavebreaking regime. However, the output pulse spectrum in that
experiment consists of several peaks \cite{IAP_exp,Kartashov_privat} similar to the observed spectrum in the experiment Ref.~\cite{Dreher}
performed in the superradiant regime \cite{superradiant}. In this regime the bounce
frequency of the laser beatwave ponderomotive potential exceeds the plasma frequency.   Thus, an alternative explanation is that the superradiant regime may have been reached in the  IAP experiments because of the low plasma density,
$n\sim(1.4-4)\times10^{16}$ cm$^{-3}$, compared to other BRA experiments.

To describe more accurately the plasma wavebreaking, PIC code simulations 
have been employed, although  comprehensive modeling including wavebreaking is difficult since many plasma wavelengths should be taken into account \cite{Clark_PIC,PIC_Nature}.
This task becomes even more challenging \cite{Wang_PIC} for simulating warm plasma
due to a large increase of particle number (on the order of $\exp(-2/k_p^2\lambda_D^2)$) needed
to resolve the electron distribution function in the resonant region.

\section{Pump intensity window for BRA}
\label{window}
\begin{figure}[ht]
\includegraphics[width=3.4in,keepaspectratio]{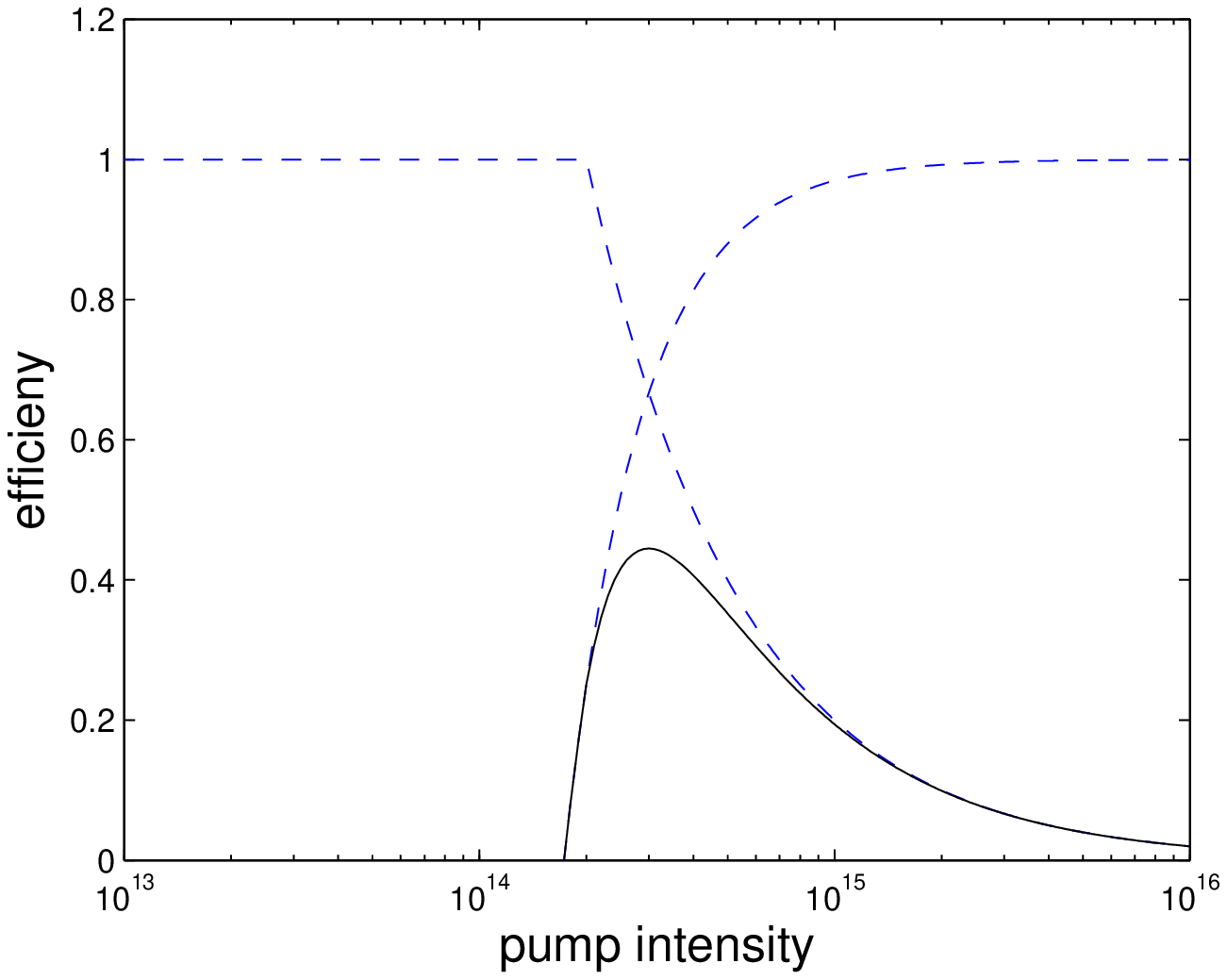}
\includegraphics[width=3.4in,keepaspectratio]{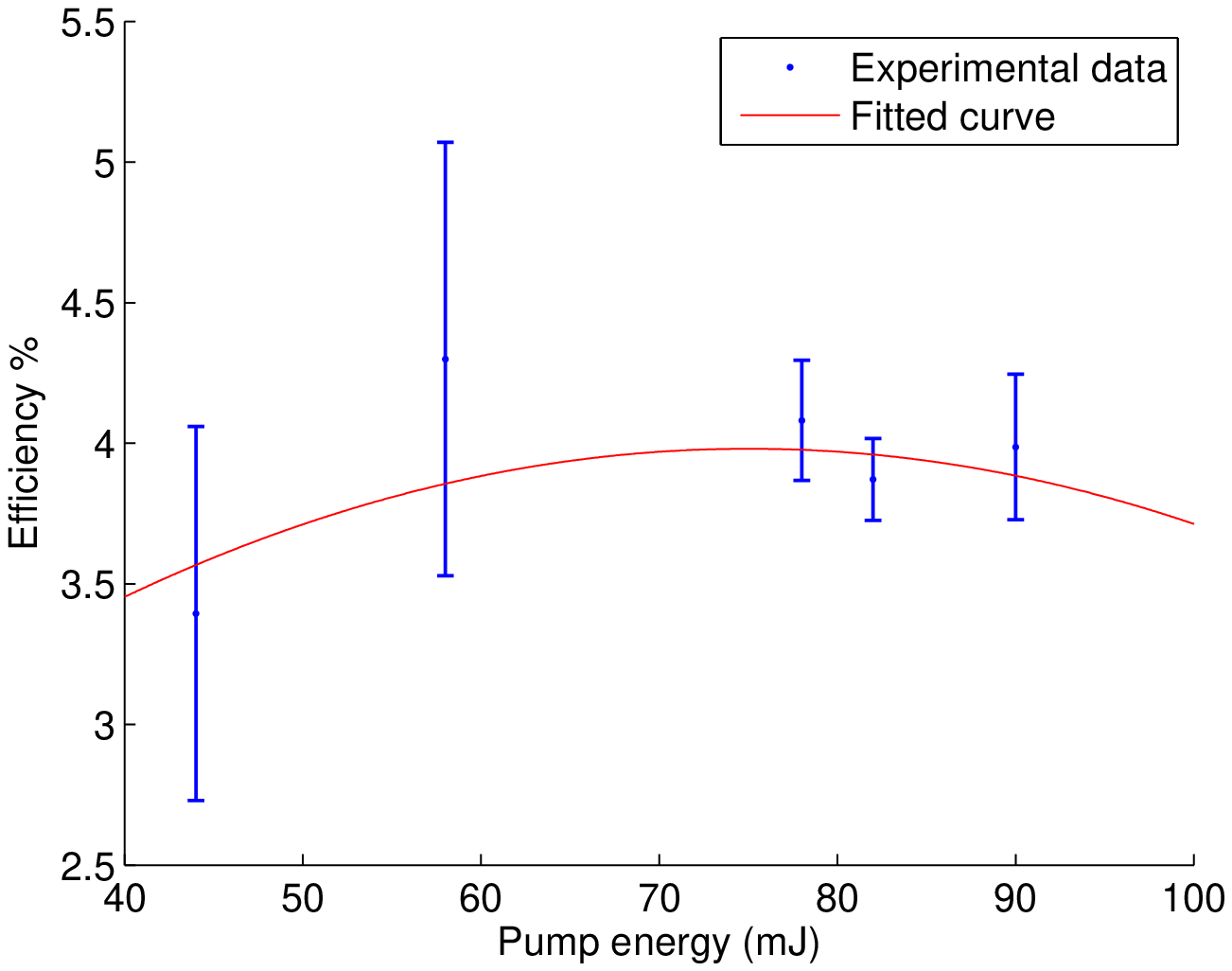}
\caption{(Color online)
The upper plot shows theoretically estimated BRA efficiency as described in the text.
The lower plot shows experimentally measured dependence of the BRA efficiency versus the pump energy in Princeton experiment \cite{Suckewer07,demonstration}.
The intensity of the pump is proportional to the pump energy since the other parameters of the pump are fixed. The solid
line is the least square parabolic fit of the experimental data points.}
\label{f4}
\end{figure}

As discussed in Secs.~\ref{detuning} and \ref{wavebreaking}, detuning and plasma wavebreaking
are important in current BRA experiments.
The plasma wavebreaking limits amplification at high pump intensity while
the detuning becomes important at low pump intensity. Therefore, there is a parameter region for the pump intensity which
is favorable for BRA. This idea is illustrated in Fig.~\ref{f4} (upper plot). Here we plot individual
efficiencies due to the detuning and the
plasma wavebreaking. The BRA efficiency due to the wavebreaking is taken from the assumption that only critical pump
intensity (\ref{Icrit}) can be absorbed. Further absorption of the pump leads to the plasma wavebreaking and cancels the
interaction. The pump depletion rate then is $D_{a}=I_a^{cr}/I_a$ if $I_a>I_a^{cr}$.
The pump depletion rate caused by the detuning was calculated in Ref.~\cite{PoP00} and it is described by Eq.~(\ref{depletion_detuning}).
The overall efficiency of BRA is assumed to be a product of partial efficiencies due to the detuning and the plasma wavebreaking.

As illustrated in Fig.~\ref{f4}, the overall BRA efficiency has a well-defined maximum at a certain pump intensity.
The experimentally observed efficiency in Princeton experiments \cite{Suckewer07,demonstration} has the same kind of intensity dependence as predicted theoretically (lower plot in Fig.~\ref{f4}).
 Although a detailed comparison of the experimental and the theoretical efficiency plots is complicated since the pump
has nonuniform intensity profile along the channel, the effects of detuning and wavebreaking are robust
and the dependence of the BRA efficiency versus pump energy can still be observed.
The overall efficiency is the largest in the regime in which these two limiting effects are competing with neither dominant. 
Thus, the experimental data supports the theoretical assumption that both the detuning and the plasma wavebreaking limit the BRA efficiency in the experiment. 

Using the estimates for BRA efficiencies due to detuning and wavebreaking described above, the optimum pump intensity and the pump depletion rate can be estimated as
\bea
\label{optimI}
I_a^{\rm opt}&=&\left\{
\begin{array}{lcl}
	I_a^{cr}\ds\frac{\sqrt{3}q_{cr}\xi_*}{4}&, &q_{cr}>\ds\frac{4}{\sqrt{3}\xi_*}\\
	I_a^{cr}&, &q_{cr}<\ds\frac{4}{\sqrt{3}\xi_*}\\
\end{array}\right.\\
\label{Da}
D_{a}^{\rm opt}&=&\left\{
\begin{array}{lcl}
	\ds\frac{8}{3\sqrt{3}}\frac{1}{q_{cr}\xi_*}&, &q_{cr}>\ds\frac{4}{\sqrt{3}\xi_*}\\
	\ds 1-\frac{q_{cr}^2\xi_*^2}{16}&, &q_{cr}<\ds\frac{4}{\sqrt{3}\xi_*},\\
\end{array}\right.
\eea
where $q_{cr}={2(\partial_z\omega_p-2\partial_z\omega_a)c}/({a_{0 cr}^2\omega_a\omega_p})$ is the detuning factor at critical pump intensity of the wavebreaking. Therefore, the optimum pump intensity for
the highest BRA efficiency can exceed the critical pump intensity for the plasma wavebreaking at high detuning parameters $q_{cr}>\sim0.25$ (assuming $\xi_*\sim10$ as will be discussed in
Sec.~\ref{epsilon}). At the same time, the maximum pump depletion rate remains bellow 2/3 in this regime. BRA experiments should be designed in the opposite regime in order to reach high amplification
efficiency. The optimal pump intensity in that regime is equal to the critical pump intensity for the wave breaking described by Eq.~(\ref{Icrit}) and the detuning parameter should be small.
If detuning is dominated by the pump chirp, then this regime can be achieved at small pump bandwidth
\be
\label{optimomega}
\frac{\Delta\omega_a}{\omega_a}[\%]<0.55\frac{\Delta t_a[{\rm ps}]\lambda^3[{\rm \mu m}]}{\xi_*}\left(\frac{n}{10^{19}\,{\rm cm}^{-3}}\right)^2.
\ee
Note that this condition is based on the estimate (\ref{Icrit}) for the critical pump intensity due to plasma wavebreaking in cold plasma. The critical pump intensity in experiments may be significantly
smaller due to finite plasma temperature. As a result, the critical detuning factor $q_{cr}$ 
may be larger and the requirements for the pump bandwidth are more severe.
This assumption is supported by the experimental data at Princeton experiments
\cite{Suckewer07,demonstration}. Condition (\ref{optimomega}) was roughly satisfied in that setup, however, the pump depletion rate in axial experiments
was on the order of $10\%$ in the laser coupling region. This value is well below the
theoretical estimate (\ref{Da}) which indicates that critical pump intensity due to wavebreaking was significantly
reduced in warm plasma. At the same time, the critical detuning parameter $q_{cr}$ was reduced in the tilted laser pulse experiments, which resulted in the pump depletion rate (about $40\%$) significantly
closer to the analytical estimate.

\section{Landau damping}
\label{Landau}

As discussed in Sec.~\ref{wavebreaking}, plasma wavebreaking in BRA occurs when the plasma wave amplitude exceeds
the critical value. 
However, in
warm plasma there might be significant number of electrons interacting with the plasma wave even at small wave amplitudes
resulting in Landau damping of the plasma wave. 
As a result, the laser coupling is reduced, diminishing the BRA efficiency.

 A similar  Landau damping effect occurs in
inertial confinement fusion (ICF) plasma \cite{ICF1,ICF2,ICF3}. The plasma wave both in BRA and ICF is generated by the backward Raman 
scattering (BRS). Studies of Landau damping in ICF plasma are motivated by the suggestion that the BRS can be strongly suppressed
at large values of $k_p\lambda_D$.  As opposed to the case of ICF, Landau damping in the case of BRA is unwanted since it reduces laser coupling and, therefore, the amplifier
should be operated in the regime of small $k_p\lambda_D$.
Whether wanted or not, for both applications it is important to estimate accurately the Landau damping, which is complicated by the presence of the pump waves.
The inverse bremsstrahlung losses of the pump  can increase the plasma temperature
leading to the regime of high $k_p\lambda_D$.  On the other hand, the
electron distribution function might be modified by a large amplitude plasma wave so that the Landau damping
might saturate \cite{Zakharov,Mazitov,ONeil,Dewar}. Therefore, these competing effects must be considered in describing Landau damping when the plasma wave is externally
driven by the ponderomotive force of the laser beatwave.

A fully kinetic description of nonlinear Landau damping is a challenging task both
numerically and analytically. A number of reduced analytical models
\cite{Benisti07,Lindberg08,NL_Landau} as well as simplified codes 
\cite{Hur04,Wang_3wave,Benisti_code} describing nonlinear Landau damping of a driven plasma wave were recently offered. The quasilinear model described in Ref.~\cite{NL_Landau} was chosen to study this effect since it is formulated
in terms of simple equations which can be analyzed analytically.
This model is in a reasonable agreement with more precise analytical kinetic model developed in Refs.~\cite{Benisti07} and numerical simulations \cite{Benisti_me}.
The quasilinear model, however, does not account for a number
of important effects such as particle detrapping due to their ballistic motion across the speckle or collisions.

The analysis of nonlinear Landau damping in BRA is presented in Ref.~\cite{BRA_Landau}. 
A regime was found in which the initially large linear Landau damping rate can become saturated
The main scaling in this estimate comes from
the exponential dependence of the linear Landau damping rate on the plasma temperature. Therefore, the estimate for the maximum
plasma temperature in BRA is roughly the same for various experiments described above \cite{Suckewer07,LLNL09,Tai_exp}.
\be
\label{kD}
k\lambda_D<0.4,\;\;\;\;\;\;\;\;\;T<190\left(\frac{n}{10^{19} {\rm cm^{-3}}}\right)\lambda^2[\rm \mu m]\; {\rm eV},
\ee
for $I_a\sim10^{14}{\rm W/cm^2}$, $n\sim10^{19}{\rm cm^{-3}}$, and few milimeter plasma length.

Estimate (\ref{kD}) roughly agrees with findings presented in Ref.~\cite{Wang_PIC}, where no significant amplification was observed at high plasma temperature,
$k_p\lambda_D>0.4$. However, parameter region for efficient BRA presented in Ref.~\cite{Wang_PIC} was strongly affected by the forward Raman scattering (FRS)
of the amplified pulse. The growth rate of FRS is about $(\omega_p/\omega)^2$ times smaller than the growth rate of BRS. Therefore, FRS
instability is more sensitive to plasma density fluctuations than BRS. Current experimental data indicate noticeable plasma density fluctuations \cite{LLNL07,Suckewer07} which 
are likely to suppress FRS \cite{Tsidulko_scat} in modern BRA experiments unlike analytical and numerical predictions which assume uniform plasma \cite{prl99,PoP00,Wang_PIC,PIC_Nature}

Estimate (\ref{kD}) suggests that the Landau damping does not affect strongly the amplification in the Princeton experiments
\cite{Suckewer07} ($k_p\lambda_D\approx 0.23$).
This conclusion is also supported by experiments with tilted geometry of laser pulses  in which the detuning was
compensated on the side of the pump entrance into the channel. The plasma is likely the hottest in that domain since it was heated for the longest time.
In that case, if Landau damping were strong,  the detuning compensation might not be effective.

On the other hand, the estimated plasma temperature in the LLNL experiments ($T=200-250$ eV \cite{LLNL09}, $k_p\lambda_D=0.38-0.42$,) suggests that Landau
damping probably played a more significant role in those experiments, which is confirmed by 2D PIC
simulations \cite{LLNL11}. The same estimate for the parameters of experiment performed
at IAMS ($T=150-200$ eV \cite{Tai_exp}, $k_p\lambda_D=0.55-0.65$) suggests that that Landau damping was the dominant limiting effect
in that experiment. It would appear, however, that the large amplification of the seed pulse
(\ref{kD}) suggests either a much lower temperature or that the BRS instability is not suppressed in hot plasma.
The latter possibility would have  important implications for the NIF program.   Accurate conclusions about 
Landau damping in LLNL and IAMS experiments, however, can only be drawn with more accurate
measurements of the plasma temperature. Measuring the plasma temperature, however,  can be challenging since the plasma temperature changes during amplification due to
inverse bremsstrahlung losses of the pump, so that  the plasma temperature during
laser amplification may be unrelated to the plasma temperature before interaction.

Note that plasma heating can be avoided in yet unpursued ``ionization front" regime wich can be achieved if a short intense seed
pulse ionizes plasma and seeds amplification at the same time \cite{self_ion,self_ion_num}. As a result, the pump does not heat plasma for a long
time prior to laser coupling and Landau damping is expected to be small in this regime.

\section{Effective seeding power}
\label{epsilon}

Estimates for the pump depletion rate in the presence of detuning (Eqs.~(\ref{depletion_detuning}), (\ref{optimI})---(\ref{optimomega}) ) as well as estimates
for saturation of Landau damping in BRA \cite{BRA_Landau} depend on the
self-similar position of the amplified pulse maximum $\xi_*$. This coordinate weakly depends on the integrated amplitude of the seed pulse
$\epsilon=\int b(\tau=0,\zeta) d\zeta$
\be
\label{eps}
\xi_*=\ln(4\sqrt{2\pi\xi_*}/\epsilon),
\ee
\be
\epsilon=0.18\sqrt{\frac{I_b\;\lambda[{\rm \mu m}]}{10^{12}\;{\rm W/cm^2}}}\left(\frac{n}{10^{19}\;{\rm cm^{-3}}}\right)^{1/4}\Delta t_b[{\rm ps}]
\ee
where $I_b$ and $\Delta t_b$ are the seed pulse intensity and FWHM duration, respectively.

The integrated seed pulse amplitude $\epsilon$ is equal to the pump depletion rate behind the seed pulse in the linear regime. This parameter fully describes the
evolution of the amplified pulse within the three-wave coupling model (\ref{eqa_reduced}) under assumption that the seed pulse is infinitely
short. The integrated amplitude of the seed pulse in all the experiments is large enough ($\epsilon\sim0.1$, $\xi_*\approx5.5$
in Princeton experiments) to provide significant amplification. However, the integrated
seed pulse amplitude cannot be used to characterize the amplified pulse growth if the seed pulse is long since only a small fraction of the seed pulse contributes
to BRA seeding in this regime. Qualitatively, the effect of the finite seed pulse duration plays role when the duration of the amplified pulse becomes on the order or smaller then the
duration of the seed pulse. In this regime the laser coupling region is located at the front of the seed pulse and the back part of the seed pulse
remains behind the interaction region. Autocorrelation measurements of the original seed and the amplified pulse durations
in Princeton experiments \cite{Suckewer07} show that the amplified pulse becomes much shorter than the original seed pulse. The effective seeding power which contributes to
the pulse amplification is significantly reduced in this regime.

The quantitative estimate for the effective seeding power was presented in Ref.~\cite{epsilon_eff} assuming Gaussian temporal profile of the seed.
It was shown that the domain which contributes the most to the pulse amplification slips to the front of the seed pulse and its duration decreases during
amplification. Eventually, only a small domain with exponentially small amplitude effectively drives amplification. The quantitative estimate for effective seeding power
reads as \cite{epsilon_eff}
\be
\label{xi_eff}
\xi_{\rm eff}\approx\xi_*+\frac{304}{\xi_{\rm eff}^2}\left(\frac{I_a L[{\rm mm}] \Delta t_b[{\rm ps}]}{10^{14} {\rm W/cm^2}}\right)^2\frac{n \lambda^2[{\rm \mu m}]}{10^{19}{\rm cm^{-3}}},
\ee
where $L$ is the plasma length. 

The regime of strongly reduced effective seeding power is likely to take place in experiments performed at Princeton \cite{Suckewer07}. The analytical estimate (\ref{xi_eff})
for the effective seeding power predicts the increase of $\xi_{\rm eff}$ by more than a factor of 2. Note that the effective seeding power described by Eq.~{\ref{eps}}
depends exponentially on the effective self-similar coordinate of the 
amplified pulse maximum $\xi_{\rm eff}$. Therefore, the effective seeding power should decrease by several orders of magnitude.
In this regime, the effective seeding domain is located on the front tail of the original seed pulse and the estimate (\ref{xi_eff}) cannot be used since small
deviation of the seed pulse from the Gaussian profile causes large discrepancy with analytical model.
Poor pulse profile can even result in appearance of precursors \cite{precursor} which prevent pulse compression
and growth of its peak intensity.

Currently, the seed pulses are generated by downshifting the frequency of the chirped pulse in a Raman cell \cite{Suckewer07,LLNL07,LLNL09} or
through generation of supercontinuum \cite{Tai_exp} with the following temporal compression. Such pulses are expected to have pure shot-to-shot stability and poor front quality and,
thus, are subject to generate precursors. This problem can be avoided upon increasing the seed pulse contrast using well-developed techniques. It is also worth noting that poor contrast of
the seed pulses in current experiments does not result in a decrease of the effective seeding
power by many orders of magnitude. Therefore, it is reasonable to consider $\xi_{\rm eff}\sim10$
in all the estimates presented in previous sections.

\section{Discussion}

It is possible to deduce in  recent state-of-the-art experiments that  the experimentally
realized efficiency is likely constrained mainly by  the pump chirp and wavebreaking, which can be put as a
bracket on pump intensity.    The plasma density gradient was shown to compensate for the pump chirp. 
When these explanations are taken into account, together with an accounting of the spatial and temporal overlap
of the counterpropagating beams, it becomes clear that the experimentally realized efficiency is within the theoretical expectations.
In addition, the plasma temperature should be small enough, $k_p\lambda_D<0.4$, to avoid Landau damping of the plasma
wave which mediates the laser coupling.


Moreover, although the experiments were carried out in the optical regime for ps pulses, it is clear that the
establishment of the physical mechanisms at play here can be extrapolated to other regimes of resonant Raman 
compression, such as through ionization-front coupling  \cite{self_ion}, X-ray or UV compression   
\cite{Malkin_ultra, Malkin_atto, Malkin_relic, Son_xray1, Son_xray2} , or the quasitransient regime  \cite{Malkin_quasi1, Malkin_quasi2}.


\end{document}